\documentclass[11pt,superscriptaddress,aps,prd,preprint]{revtex4}
\usepackage{amsmath}

\makeatletter
\usepackage[T1]{fontenc}
\usepackage{amsmath}
\usepackage{amssymb}
\usepackage{graphicx}

\usepackage{slashed}
\usepackage{xcolor}

\newcommand{\bea}{\begin{eqnarray}}
\newcommand{\eea}{\end{eqnarray}}

\begin{document}

\title{G\"{o}del-type universes in bumblebee gravity}

\author{W. D. R. Jesus}\email[]{willian.xb@gmail.com}
\affiliation{Instituto de F\'{\i}sica, Universidade Federal de Mato Grosso,\\
78060-900, Cuiab\'{a}, Mato Grosso, Brazil}

\author{A. F. Santos}\email[]{alesandroferreira@fisica.ufmt.br}
\affiliation{Instituto de F\'{\i}sica, Universidade Federal de Mato Grosso,\\
78060-900, Cuiab\'{a}, Mato Grosso, Brazil}

\begin{abstract}

The bumblebee field coupled with gravity is considered. This gravitational theory exhibits spontaneous breaking of Lorentz symmetry. The G\"{o}del-type universe is introduced and then the causality and its violation are studied. Causal and non-causal G\"{o}del-type solutions are obtained for different content of matter. In addition, when the coupling constant, which controls the interaction between the bumblebee field and the gravitational field, is zero the bumblebee potential may be associated with the value of the cosmological constant of the general relativity. Furthermore, the case with the non-zero coupling constant is also investigated.

\end{abstract}

\maketitle

\section{Introduction}

Standard model (SM) and General Relativity (GR) are examples of theories that describe the fundamental forces of nature. These theories have been extensively tested over the years and their theoretical predictions are confirmed experimentally. Although SM describes fundamental interactions at a quantum level, GR is a classical theory. So, at the present moment, a consistent quantum gravity theory, which unifies GR and quantum mechanics, has not been constructed. However, some models that unify SM and GR have been developed. A unified theory leads to a deeper understanding of nature. The physical effects of this unification theory must emerge at very high energies, i.e., at the Planck scale ($\sim 10^{19}\mathrm{GeV}$). On this scale of energy, a new physics arises. For example, the Lorentz and CPT symmetries that are the foundation of SM and GR may be broken \cite{KS}. To study these ideas, an extension of SM has been developed.

The Standard Model Extension (SME) \cite{SME1, SME2} is an effective field theory that describes the Lorentz violation, both theoretical and experimental. The SME is composed of known physics of the SM plus all possible terms that violate Lorentz and CPT symmetry. The GR in the framework of the SME has been considered \cite{Kos0, Kos01, Kos02}. The gravitational sector of the SME contains 19 coefficients for Lorentz violation and an unobservable scalar parameter. This model was developed using the ideas of spontaneous Lorentz symmetry breaking. Here the bumblebee model is considered. It is the simplest example of a theory with spontaneous Lorentz symmetry breaking, i.e.,  a vector field acquires a nonzero vacuum expectation value implying that the vacuum of the theory gets a preferential direction in the space-time. The bumblebee model has been extensively studied in the literature \cite{Ovgun:2018xys, Casana:2017jkc, Capelo:2015ipa, Nascimento:2014vva, Her, Maluf:2015hda}. In this paper, the question of causality that emerges from the G\"{o}del-type universe in the bumblebee model will be investigated.

The violation of causality in GR is a permissible phenomenon since exact solutions in this theory allow the presence of Closed Time-like Curves (CTC's). There are some models that allow the existence of CTC's, as a consequence a traveler who moves along such curves can return to his own past. Examples that exhibit such characteristic are, the G\"{o}del universe \cite{Godel49}, the van Stockum space-time of a rotating dust cylinder \cite{Van}, the Kerr space-time \cite{Kerr} and the Gott space-time of two cosmic strings \cite{Gott}, among others. A review of cosmological models with rotation that exhibit causal violation is presented in \cite{Obu}. Here the G\"{o}del-type solutions are considered. This solution was proposed by K. G\"{o}del. It is an exact solution of GR with rotating matter. In addition, it is stationary,
spatially homogeneous and possessing cylindrical symmetry. A generalization of the G\"{o}del solution has been developed, such solutions are denominated by G\"{o}del-type metrics \cite{PRD28}. In these metrics the question of causality is investigated in details and there is possibility of eliminating the CTC's for determined values
of their parameters. Furthermore, with such generalization three different classes of solution may be analyzed: (i) there is no CTC's, (ii) there is an infinite sequence of alternating causal and non-causal regions, and (iii) there is only one non-causal region. The G\"{o}del-type solution has been studied in several modified gravity theories \cite{CS, CS1, fR, fR1, fT, fRT, fRT1, Hor, Bum, BD, fRQ}. Here causal and non-causal regions in the bumblebee gravity are determined.

This paper is organized as follows. In section II, a brief introduction to bumblebee model is presented. In section III, the  G\"{o}del-type universes are discussed. In section IV, the G\"{o}del-type universe in bumblebee theory, for different matter contents, is studied. The analysis considers two cases, that is, with the coupling constant zero and non-zero. Causal and non-causal G\"{o}del-type solutions are found.  In section V, some concluding remarks are discussed.

\section{Bumblebee Model}

Here the Bumblebee model is presented. This theory considers a vector field $B_\mu$ which acquires a nonzero vacuum expectation value $b_\mu$ that  allows spontaneous Lorentz symmetry breaking. The action that captures the most relevant features of the bumblebee model describes the coupling between the bumblebee field and geometry and the presence of a general potential. A discussion of the symmetries of a model with couplings of the form $B^\mu B^\nu R_{\mu \nu}$ and $B^\mu B_\mu R$ has been presented \cite{Capelo:2015ipa, Bluhm2}. It is interesting to note that the choice of the potential can have implications for the parameters of the theory. For example, a general potential implies that the inclusion of both types of couplings should have physical consequences. Here, for simplicity, the coupling to the Ricci scalar is discarded. 
Thus the model studied here is the same as presented and investigated in Ref. \cite{Kos0}.

The action that describes the dynamics of the bumblebee field is given by \cite{KS, Kos0, Kos01, Bluhm2}
\bea\label{action}
S&=&\int \sqrt{-g}\Bigl[\frac{1}{2\kappa}(R+\xi B^\mu B^\nu R_{\mu \nu}) -\frac{1}{4}B^{\mu \nu}B_{\mu \nu}-V(B^\mu B_\mu \pm b^2) +\mathcal{L}_M \Bigl],
\eea
where $g$ is the metric determinant, $\kappa=8\pi G$, $\xi$ is the coupling constant and $\mathcal{L}_M$ is the Lagrangian density for the matter fields. The field-strength tensor $B_{\mu\nu}$ is defined as
\bea
B_{\mu \nu}\equiv \partial_\mu B_\nu-\partial_\nu B_\mu
\eea
and the potential $V(B^\mu B_\mu \pm b^2)$ is responsible for triggering the mechanism of spontaneous Lorentz violation. The field $B_\mu$ takes on a nonzero vacuum value, i.e. $\langle B_\mu\rangle=b_\mu$. Then for a local minimum at  $B_\mu B^\mu \pm b^2=0$ that leads to $b^\mu b_\mu=\mp b^2$. Therefore, the non-null vector background $b_\mu$ spontaneously breaks the Lorentz symmetry. A correspondence between the bumblebee action (\ref{action}) and the Lorentz-violating action obtained in the gravitational sector of minimal SME has been established \cite{Kos02}. The action of the gravitational sector provided by the minimal SME is given as
\bea
S_{LV}=\frac{1}{2\kappa}\int d^4x\sqrt{-g}\left[uR+s^{\mu\nu}R_{\mu\nu}+t^{\mu\nu\alpha\rho}R_{\mu\nu\alpha\rho}\right],
\eea
where $u$, $s^{\mu\nu}$ and $t^{\mu\nu\alpha\rho}$ are real and dimensionless tensors that carry information about Lorentz violation. The correspondence between the bumblebee and SME action is 
\bea
u=\frac{1}{4}\xi B^\mu B_\mu, \quad s^{\mu\nu}=\xi\left(B^\mu B^\nu-\frac{1}{4}g^{\mu\nu}B^\alpha B_\alpha\right), \quad t^{\mu\nu\alpha\rho}=0.
\eea

By taking the action (\ref{action}) and varying it with respect to the metric tensor $g_{\mu \nu}$, the modified Einstein equation is given as
{\small
\bea
G_{\mu \nu}&=&\kappa \left[2V'B_\mu B_\nu+B_{\mu \alpha}B_\nu{}^\alpha-\left(V+\frac{1}{4}B_{\alpha \beta}B^{\alpha \beta}\right)g_{\mu \nu} \right]+\xi \Biggl[\frac{1}{2}B^\alpha B^\beta R_{\alpha \beta}g_{\mu \nu}-B_\mu B^\alpha R_{\alpha \nu}-B_\nu B^\alpha R_{\alpha \mu}\nonumber\\ 
&+&\frac{1}{2}\nabla_\alpha \nabla_\mu(B^\alpha B_\nu)+\frac{1}{2}\nabla_\alpha \nabla_\nu(B^\alpha B_\mu)-\frac{1}{2}\nabla_\alpha \nabla_\beta(B^\alpha B^\beta)g_{\mu \nu}-\frac{1}{2}\square(B_\mu B_\nu)\Biggl]+\kappa\, T_{\mu \nu},
\eea}
where $G_{\mu \nu}$ is the Einstein tensor, $V'$ denotes the derivative of the potential $V$ with respect to its argument and $T_{\mu \nu}$ is the energy-momentum tensor of matter that is defined as
\bea
T_{\mu\nu} = -\frac{2}{\sqrt{-g}}  \frac{\delta \left(\sqrt{-g}{\cal L}_M\right)}{\delta g^{\mu\nu}}.
\eea

Similarly, varying the eq. (\ref{action}) with respect to the field $B_\mu$, the equation of motion for the bumblebee field is given as
\begin{equation}\label{eqmov}
\nabla_\mu B^{\mu \nu}=2\left(V'B^\nu-\frac{\xi}{2\kappa}B_\mu R^{\mu \nu}\right).
\end{equation}
Note that when bumblebee field $B_\mu$ and potential $V(B_\mu)$ are null, the original general relativity field equations are recovered.

In the next section, the G\"{o}del-type solutions are presented. Then the consistency of these solutions in the bumblebee gravity theory will be verified.

\section{G\"{o}del-type Universes}

In this section, a brief introduction to the G\"{o}del-type universes is presented.  The G\"{o}del metric was constructed by Kurt G\"{o}del, in 1949 \cite{Godel49}. It is an exact solution with rotating matter of the Einstein field equations. The main feature of this metric is the possibility of Closed Time-like Curves (CTC's), which lead to the violation of causality. In order to study the question of causality with more details, an extension of this solution has been proposed \cite{PRD28}. This generalization is known as G\"{o}del-type solution and it is described by the line element
\begin{equation}\label{eq:15}
ds^2=[dt+H(r)d\phi]^2-dr^2-D^2(r)d\phi^2-dz^2,
\end{equation}
where the functions \(H(r)\) and \(D(r)\) satisfy the relations 
\begin{equation}\label{eq:16}
\frac{H'(r)}{D(r)}=2\omega \qquad \mathrm{and} \qquad \frac{D''(r)}{D(r)}=m^2,
\end{equation}
where the line denotes derivative with respect to $r$, \(m\) and \(\omega\) are constant parameters. These parameters are used to define different G\"{o}del-type geometries, such that \(\omega >0\) and \(-\infty\leq m^2 \geq +\infty\). The G\"{o}del-type solution, eq. (\ref{eq:15}), can be written as
\begin{equation}\label{eq:17}
ds^2=dt^2+2H(r)dtd\phi-dr^2-G(r)d\phi^2-dz^2,
\end{equation}
where $G(r)=D^2(r)-H^2(r)$. The circles defined by $t,z,r=constant$, that depend on the behavior of the $G(r)$, lead to the existence of CTC's when $G(r)<0$ for a certain range of $r$.

There are three different classes of G\"{o}del-type solutions: (i) linear, when $m=0$ and there are non-causal G\"{o}del circles; (ii) trigonometric, for $m^2 \equiv \mu ^2<0$ with an infinite sequence, alternating between causal and non-causal regions and (iii) hyperbolic, for $m^2>0$. There is one non-causal region. In addition,  causal solutions are possible too. Here the hyperbolic class is considered. In this case, the functions $H(r)$ and $D(r)$ are defined as
\begin{equation}\label{eq:110}
H(r)=\frac{4\omega}{m^2}sinh^2\left(\frac{mr}{2}\right) \quad \mathrm{and} \quad D(r)=\frac{1}{m}sinh(mr),
\end{equation}
and the non-causal regions occur for $r > r_c$ such that
\begin{equation}\label{eq:111}
sinh^2\left(\frac{mr_c}{2}\right)>\left(\frac{4\omega^2}{m^2}-1\right)^{-1},
\end{equation}
with $r_c$ being the critical radius.
Here two situations can be analyzed: (i) If $m^2=2\omega^2$ the G\"{o}del solution \cite{Godel49} is recovered and the critical radius, beyond which the causality is violated, is
\begin{equation}\label{eq:112}
r_c=\frac{2}{m}sinh^{-1}(1).
\end{equation}
(ii) If $m^2=4\omega^2$ the critical radius $r_c \rightarrow \infty$, then the violation of causality is avoided.

In order to facilitate the calculation of the field equations, let us use the following basis
\begin{equation}\label{eq:114}
\theta^0=dt+H(r)d\phi, \quad \theta^1=dr, \quad \theta^2=D(r)d\phi, \quad \theta^3=dz,
\end{equation}
where the line element takes the form
\begin{equation}\label{eq:113}
ds^2=\eta_{ab}\theta ^a\theta ^b=(\theta^0)^2-(\theta^1)^2-(\theta^2)^2-(\theta^3)^2,
\end{equation}
with \(\theta^a=e^a{}_\mu dx^\mu\) being the 1-form and $\eta_{ab}=diag(+1,-1,-1,-1)$ being the Minkowski metric. In this new basis, a flat space-time, the non-zero Ricci tensor components are given by
\bea\label{eq:115}
R_{00}&=&\frac{1}{2}\left(\frac{H'}{D}\right)^2=2\omega ^2,\\
R_{11}&=&R_{22}=R_{00}-\frac{D''}{D}=2\omega^2-m^2,
\eea
and the scalar curvature is $R=2(m^2-\omega^2)$. Then the non-zero Einstein tensor components become
\begin{equation}\label{eq:117}
G_{00}=3\omega^2-m^2, \qquad G_{11}=G_{22}=\omega^2, \qquad G_{33}=m^2-\omega^2.
\end{equation}
These results will be used in the next section, where the G\"{o}del-type solution is studied in the bumblebee gravity framework.

\section{G\"{o}del-type Solution in Bumblebee Gravity}

Here the G\"{o}del-type solution in bumblebee gravity is analyzed. The possibility of causality violation in this framework is investigated. The field equations of the bumblebee model are written in the tetrad frame as
\bea
    G_{ac}&=&\kappa\left[2V'B_aB_c-B_{ad}B_c{}^d-\left(V+\frac{1}{2}B_{de}B^{de}\right)\eta_{ac}\right]+\xi\Biggl[\frac{1}{2}B^dB^eR_{de}\eta_{ac}-B_aB^dR_{dc}-B_cB^dR_{da}\nonumber\\
    &+&\frac{1}{2}\nabla_d\nabla_a(B^dB_c)+\frac{1}{2}\nabla_d\nabla_c(B^dB_a)-\frac{1}{2}\nabla_d\nabla_e(B^dB^e)\eta_{ac}-\frac{1}{2}\square(B_aB_c)\Biggl]+\kappa\, T_{ac}.\label{fe}
\eea

In order to calculate the field equations, an ansatz for the bumblebee field is chosen as
\begin{equation}
    B_a=(B(t),0,0,0),
\end{equation}
which satisfies the condition $\eta ^{ac}B_aB_c=\pm b^2$.

Due to the difficulty of solving the field equations, let's restrict ourselves to two different cases. In the first case, the coupling constant $\xi$ is equal to zero. And in the second case, the vacuum solution is considered, where the bumblebee field becomes a constant that minimizes the potential.

\subsection{First case: $\xi=0$}

In this case, where the coupling constant vanishes ($\xi=0$), the field equation associated with the bumblebee field $B_a$, eq.~(\ref{eqmov}), give us
\begin{equation}
    V'B=0. 
\end{equation}
This forces the bumblebee field to rest at one of the extremes of its potential and keeps it from evolving with time. Thus, setting $\xi=0$, we get $V'=0$, $\Dot{B}=\Ddot{B}=0$ and $V=V_0$, with $V_0$ being a constant non zero. Then the field equations, eq. (\ref{fe}), become
\begin{equation}
    G_{ac}=-\kappa V_0\eta_{ac}+\kappa T_{ac}.
\end{equation}
Therefore, the bumblebee field contributes only through constant potential. In order to solve this field equation, a perfect fluid is considered as matter source. Its energy-momentum tensor in the tetrad basis is given as
\begin{equation}
    T^{(M)}_{ab}=(\rho+p)u_au_b-p\eta_{ab},\label{PF}
\end{equation}
where $\rho$ is the density energy, $p$ is the fluid pressure and $u_a=(1,0,0,0)$ is the 4-velocity. The non-zero components are
\begin{equation}
    T^{(M)}_{00}=\rho, \qquad T^{(M)}_{11}=T^{(M)}_{22}=T^{(M)}_{33}=p.
\end{equation}
Then, the non-null components of the field equations are
\bea
    3\omega^2-m^2&=&\rho-V_0\label{A1},\\
    \omega^2&=&p+V_0,\label{A2}\\
   m^2-\omega^2&=&p+V_0.\label{A3}
\eea
Combining these equations, we obtain
\bea
 m^2 &=&2\omega^2,\label{S1godel1}\\
    \rho &=&\omega^2+V_0,\label{S1godel2}\\
    p &=&\omega^2-V_0.\label{S1godel3}
\eea
The eq.~(\ref{S1godel1}) corresponds to the standard G\"{o}del solution that leads to causality violation. 
The equations (\ref{S1godel1})-(\ref{S1godel3}) have a close resemblance to equations of general relativity with $-V_0$ playing the role of the cosmological constant.

Here other sources of matter are considered. The main objective is to investigate whether other  sources of matter can generate causal solutions. First, let's consider two different sources of matter, namely a combination of a perfect fluid with a scalar field. Then the total energy-momentum tensor is given as
\bea
T_{ab}= T^{(M)}_{ab}+T^{(S)}_{ab},\label{PF+SF}
\eea
where $T^{(M)}_{ab}$ is the energy-momentum tensor of the perfect fluid, eq. (\ref{PF}), and $T^{(S)}_{ab}$ is the energy-momentum tensor of the scalar field that is defined as
\begin{equation}\label{SF}
T^{(S)}_{ab}=\nabla_a\Phi\nabla_b\Phi-\frac{1}{2}\eta_{ab}\eta^{cd}\nabla_c\Phi\nabla_d\Phi,
\end{equation}
with \(\nabla_a\) being the covariant derivative with respect to base \(\theta^a=e^a{}_bdx^b\). Assuming, for simplicity, that
\begin{equation}
\Phi(z)=sz+c_1,
\end{equation}
where $s$ and $c_1$ are constants, the non-vanishing components of energy-momentum tensor for this scalar field are
\begin{equation}
T^{(S)}_{00}=-T^{(S)}_{11}=-T^{(S)}_{22}=T^{(S)}_{33}=\frac{s^2}{2}.
\end{equation}

For this source of matter, the field equations become
\bea
    3\omega^2-m^2&=&\rho-V_0+\frac{s^2}{2},\\
    \omega^2&=&p+V_0-\frac{s^2}{2},\\
   m^2-\omega^2&=&p+V_0+\frac{s^2}{2}.
\eea
Combining these equations, we obtain
\bea
  &&\rho =\omega^2+V_0-\frac{3}{2}s^2,\label{FS1}\\ 
   &&p =\omega^2-V_0+\frac{s^2}{2},\label{FS2}\\ 
    &&m^2-2\omega^2=s^2.\label{FS3}
\eea
It is important to note that, due to the introduction of the scalar field, the freedom to choose the $m^2$ value emerges. If $s^2 > 0$, the condition $m^2 = 4\omega^2$ is a possible value which leads to a causal G\"{o}del-type solution.

The positive value of the density $\rho$ and the pressure $p$ are insured if the value of the constant potential is bounded within the interval
\begin{equation}
  \frac{3}{2}s^2-\omega^2\leq V_0\leq \frac{1}{2}s^2+\omega^2,  
\end{equation}
that leads to
\begin{equation}
    2\omega ^2\geq s^2, \label{w-s1}
\end{equation}
which implies that
\begin{equation}
    V_0=2\omega^2.\label{V-s1}
\end{equation}
These values lead to a causal solution, which is allowed to $m^2=4\omega^2$. And from the equations (\ref{FS1}) and (\ref{FS2}), we get $\rho = p = 0$.  An important note, by comparing the analysis done here with the discussion realized in \cite{PRD28}, the potential $V_0$ may be interpreted as being the negative of the cosmological constant, i.e., $V_0=-\Lambda$.

Now, an electromagnetic field aligned on $z$-axis and dependent of $z$ is added to our matter content. The non-vanishing components of the electromagnetic tensor in tetrad frame are
\begin{equation}
  F_{03}=-F_{30}=E(z), \quad\quad F_{12}=-F_{21}=B(z),
  \label{rot}
  \end{equation}
with the solutions of the Maxwell equations given by
\begin{equation}
\begin{split}
  &E(z)=E_{0}\cos[2\omega(z-z_{0})],\\
  &B(z)=E_{0}\sin[2\omega(z-z_{0})],
  \end{split}
\end{equation}
where $E_{0}$ is the amplitude of the electric and magnetic fields and $z_0$ is a constant. The non-zero components of the electromagnetic energy-momentum tensor are
  \begin{equation}
  T^{(\mathrm{EM})}_{00}=T^{(\mathrm{EM})}_{11}=T^{(\mathrm{EM})}_{22}=\frac{E_{0}^{2}}{2}, \quad\quad T^{(\mathrm{EM})}_{33}=-\frac{E_{0}^{2}}{2}.
  \end{equation}
Then the new energy-momentum tensor is given by
\bea
T_{ab}= T^{(M)}_{ab}+T^{(S)}_{ab}+T^{(EM)}_{ab}.\label{PF+SF+EM}
\eea
Thus the field equations become
\begin{align}
    & \rho =\omega^2+\frac{e^2}{2}+V_0-\frac{3}{2}s^2,\\
    & p =\omega^2-V_0-\frac{e^2}{2}+\frac{s^2}{2},\\
    & m^2 -2\omega^2=s^2-e^2.
\end{align}
The positivity of the density $\rho$ and of the pressure $p$ imply that
\begin{equation}
\frac{3}{2}s^2-\frac{e^2}{2}-\omega^2\leq V_0\leq \frac{1}{2}s^2-\frac{e^2}{2}+\omega^2,
\end{equation}
which leads to the eqs. (\ref{w-s1}) and (\ref{V-s1}), and to a causal solution, with $\rho=p=0$, only if
\begin{equation}
    e^2=0.
\end{equation}
Therefore, the causal G\"{o}del-type solution is only possible in the case where the matter source is a combination of perfect fluid and scalar field.

\subsection{Second case: $\xi \neq 0$ - Vacuum solution}

Here the vacuum solution is considered. In this case the bumblebee field $B_a$ remains frozen in its vacuum expectation value $b_a$. Hence $V=V'=0$ and then 
eq. (\ref{eqmov}) becomes
\bea
B_aR^{ab}=0.\label{bR}
\eea
For the choice $B_a=(b,0,0,0)$, this equation leads to a trivial solution, i.e., $b=0$. However, if the Bumblebee field is $B_a=(0,b,0,0)$ the field equation yields
\bea
b(2\omega^2-m^2)=0\label{bR1}
\eea
that implies in the G\"{o}del solution, i.e., $m^2=2\omega^2$. The same condition is obtained for the Bumblebee field given as $B_a=(0,0,b,0)$. Therefore, the eq. (\ref{bR}) imposes only non-causal solution in this case with non-zero coupling constant.

It is important to note that the action (\ref{action}) considered only coupling between the bumblebee field and the Ricci tensor. So a natural question arises, what happens to the results above when considering the coupling between the bumblebee field and the Ricci scalar? For such analyze the action (\ref{action}) becomes
\bea
S&=&\int \sqrt{-g}\Bigl[\frac{1}{2\kappa}(R+\xi B^\mu B^\nu R_{\mu \nu}+\chi B_\mu B^\mu R) -\frac{1}{4}B^{\mu \nu}B_{\mu \nu}-V(B^\mu B_\mu \pm b^2) +\mathcal{L}_M \Bigl],
\eea
where $\chi$ is a coupling constant. Then the bumblebee field equation becomes
\bea
\nabla_\mu B^{\mu \nu}=2\left(V'B^\nu-\frac{\xi}{2\kappa}B_\mu R^{\mu \nu}-\frac{\chi}{2\kappa}B^\nu R\right).
\eea
Thus, considering the vacuum solution, a new condition arises, that is given as
\bea
\xi B_aR^{ab} +\chi B^b R=0.
\eea
Now let's look to the G\"{o}del solution. Considering $B_a=(b,0,0,0)$, the same trivial solution $(b=0)$ is obtained. In addition, for the other choices, i.e., $B_a=(0,b,0,0)$, $B_a=(0,0,b,0)$ and $B_a=(0,0,0,b)$ the relation $0<m^2<4\omega^2$ has been found. Therefore, the non-causal G\"{o}del circles also occurs in this case.  Furthermore, although the inclusion of coupling between the bumblebee field and Ricci scalar changes the constraint (\ref{bR}), the main physical result is not modified, i.e., the non-causal G\"{o}del solution is allowed in this gravitational theory.

Now let's consider the Einstein equations, eq. (\ref{fe}), that are given by
\begin{equation}
    G_{ac}=\xi\left(\frac{1}{2}B^dB^eR_{de}\eta_{ac}-B_aB^dR_{dc}-B_cB^dR_{da}\right)+T_{ac}.
\end{equation}
These equations will be analyzed for the Bumblebee field $B_a=(0,b,0,0)$. By taking the perfect fluid, eq. (\ref{PF}), as matter content, the non-zero components of the field equations are
\begin{align}
    3\omega^2-m^2 &=\frac{\xi b^2}{2}(2\omega^2-m^2)+\rho, \\
    \omega^2 &=\frac{3\xi b^2}{2}(2\omega^2-m^2)+p,\\
    \omega^2 &=-\frac{\xi b^2}{2}(2\omega^2-m^2)+p,\\
    m^2-\omega^2 &=-\frac{\xi b^2}{2}(2\omega^2-m^2)+p.
\end{align}
Using the condition given by eq. (\ref{bR1}), $m^2=2\omega^2$, we get $\rho=p=\omega^2$. So Einstein equations are consistent with a non-causal solution as required by the bumblebee field equation. Therefore, regardless of the content of the matter, a causal solution, in this case, is forbidden.

An important note, when the couplings between bumblebee field and Ricci scalar is considered, the modified Einstein equations acquire new terms and take the form
\begin{equation}
    G_{ac}=\xi\left(\frac{1}{2}B^dB^eR_{de}\eta_{ac}-B_aB^dR_{dc}-B_cB^dR_{da}\right)-\chi\left(B_dB^dG_{ac}+RB_aB_c\right)+T_{ac}.
\end{equation}
Although the additional coupling could lead to a different field equation, the overall dynamical structure of the field equations, when applied to the G\"{o}del solution, is not substantially modified, since the non-causal solution persists in this case, as previously analyzed.

\section{Conclusion}

The bumblebee model is a gravitational theory that exhibits spontaneous Lorentz symmetry breaking. The G\"{o}del-type solution is used to study the question of causality in this gravitational model. The study developed here is divided into two cases which depend on the value of the coupling constant, $\xi$. The coupling constant controls the non-minimal gravity-bumblebee interaction. The case with $\xi=0$ leads to a constant potential as being the bumblebee field contribution. The first analysis considers the perfect fluid as a source of matter. Combining the field equations, the G\"{o}del solution is obtained. Then the causality violation is allowed. By comparing our results with the GR results, the potential $V_0$ may be associated with the cosmological constant. In order to obtain a causal solution, other sources of matter are considered. Combining perfect fluid plus scalar field as a source of matter, the freedom to choose the parameters of the metric arises and then a causal G\"{o}del-type solution is obtained. In addition, the positivity of the energy density and pressure leads to the same relation between the bumblebee potential and the cosmological constant. The third source of matter is introduced. Now, the total energy-momentum tensor is composed of a perfect fluid, scalar field and electromagnetic field. For certain restrictions on the relationship between the scalar and electromagnetic fields a causal solution is possible. In the case $\xi\neq 0$, the vacuum solution is considered, so there is no contribution due to the potential. By taking the Bumblebee field equation, the G\"{o}del solution is recovered. The Einstein equations are consistent with a non-causal solution independent of the content of the matter. Then in the case with a non-zero coupling constant, a causal solution is not allowed. Therefore, when $\xi=0$ both the causal and non-causal solutions are allowed in the bumblebee model. However, for $\xi\neq 0$, there is a restriction, only a non-causal solution is allowed.

\acknowledgments
This work by A. F. S. is supported by CNPq projects 308611/2017-9 and 430194/2018-8.

\end{document}